# Impact of Free-carrier Nonlinearities on Silicon Microring-based Reservoir Computing

Bernard. J. Giron Castro[1], Christophe Peucheret[2], Darko Zibar[1] and Francesco Da Ros[1]

1. DTU Electro, 348, Ørsteds Pl., 2800 Kongens Lyngby, Denmark
2. Univ Rennes, CNRS, UMR6082 - FOTON, 22305 Lannion, France
*bjgca@dtu.dk*

*Abstract*— We quantify the impact of thermo-optic and free-carrier effects on time-delay reservoir computing using a silicon microring resonator. We identify pump power and frequency detuning ranges with NMSE less than 0.05 for the NARMA-10 task depending on the time constants of the two considered effects.

*Keywords—Reservoir Computing, Neuromorphic Photonics, Microring resonator, Optical Computing, Machine Learning.*

## I. INTRODUCTION

Reservoir computing (RC) is a promising computing paradigm for reducing the training complexity of recurrent neural networks. Time delay-based RC (TDRC) is a scheme that offers the benefits of traditional RC by relying on time multiplexing. This technique provides multiple computational nodes that require just a single nonlinear element with a time delay that enhances the connectivity between them [1]. Silicon microring resonators (MRR) have been studied for a variety of architectures in optical computing, including building blocks for photonic TDRC schemes [1-3, 5]. Recently, a TDRC scheme using a silicon MRR with optical feedback was numerically studied for solving time-series prediction tasks [3]. However, no clear relation between the performance of the RC and the physical effects within the silicon MRR has been established in the literature, with only some brief discussion initiated in [3].

In this numerical work, we study the impact of varying the time constants of free-carriers and thermo-optic (TO) effects within the silicon microring on the error performance of an MRR-based TDRC. We show that it is possible to control and extend the frequency detuning and input power regions of low time-series prediction errors.

## II. NONLINEARITIES IN SILICON MICRORING RESONATORS

We consider an incident quasi-monochromatic electric field at a frequency $\omega_p$ close to the frequency of the (cold) silicon MRR cavity resonance $\omega_0$, such that it reinforces the generation of excess carriers by two-photon absorption (TPA). The carriers are responsible for free-carrier dispersion (FCD), which, together with the TO effect due to the increase of heat in the cavity, change the waveguide refractive index and shift the resonance frequency. However, FCD blue-shifts it, while TO effect red-shifts it [6]. The generated carriers also cause free-carrier absorption (FCA), which adds to the waveguide losses. The mode amplitude of a resonator $a$, the excess free-carrier density generated via TPA ($\Delta N$) and the temperature difference with respect to the environment ($\Delta T$) are described by the following differential equations based on temporal coupled mode theory for a silicon add-drop MRR, considering electrical fields at both input ($E_{in}$) and add ($E_{add}$) ports [3-5]:

$$\frac{da(t)}{dt} = [i\delta(t) - \gamma_{tot}(t)]a(t) + i\sqrt{\frac{2}{\tau_c}}(E_{in}(t) + E_{add}(t))e^{j\omega_p t}, \quad (1)$$

$$\frac{d\Delta N(t)}{dt} = -\frac{\Delta N(t)}{\tau_{FC}} + \frac{\Gamma_{FCA}c^2\beta_{TPA}|a(t)|^4}{2\hbar\omega_p V_{FCA}^2 n_{Si}^2}, \quad (2)$$

$$\frac{d\Delta T(t)}{dt} = -\frac{\Delta T(t)}{\tau_{th}} + \frac{\Gamma_{th}P_{abs}(t)}{mc_p}|a(t)|^2. \quad (3)$$

In (1)-(3), $\delta(t)$ is the total angular frequency detuning, which comprises $\Delta\omega = \omega_p - \omega_0$, as well as the excess detuning induced by the TO and FCD effects. $\gamma_{tot}(t)$ and $P_{abs}(t)$ denote the total losses and power absorbed in the cavity. Table I provides the definitions of the previous time-dependent variables, where $dn/dN$ and $dn/dT$ are the FCD and TO coefficients of silicon, respectively. $\alpha$ is the waveguide attenuation and $\tau_c$ is the decay rate of the energy due to the coupling between the MRR and the bus waveguides. The losses due to FCA and TPA $\gamma_{TPA/FCA}$ are defined using the expressions and values of [4]. $\tau_{FC}$ is the lifetime of the carriers, $\tau_{th}$ is the decay time of the emerging heat as it diffuses to the surroundings, and $m$ is the mass of the MRR. The contribution of the Kerr effect to the frequency shift of the resonance is considered negligible when compared to the other effects [6]. $\Gamma_{FCA/TPA}$ refer to the FCA and TPA confinement factors. $n_{Si}$, $\beta_{TPA}$ and $c_p$, are the silicon's refractive index, TPA coefficient and specific heat, respectively. $V_{FCA}$ is the FCA effective volume. The values of the silicon constants, confinements factors and $V_{FCA}$ are taken from [5].

## III. RESERVOIR SIMULATION

We normalize and numerically solve (1)-(3) with a 4$^{th}$-order Runge-Kutta method [5, 6]. The simulated TDRC system (Fig. 1) includes a delay loop connected from the through to the add port to increase the memory capacity of the RC by adding a

TABLE I.    TIME-DEPENDENT SIMULATED PARAMETERS

| Parameter | Definition [3-6] |
|---|---|
| $\delta(t)$ | $\omega_p - \omega_0[1 - 1/n_{Si}(\Delta N \cdot dn/dN + \Delta T \cdot dn_{Si}/dT)]$ |
| $\gamma_{tot}(t)$ | $n_{Si}/(c\cdot\alpha) + 2/\tau_c + \gamma_{TPA} + \gamma_{FCA}$ |
| $P_{abs}(t)$ | $n_{Si}/(c\cdot\alpha) + \gamma_{TPA} + \gamma_{FCA}$ |

Acknowledgements: Villum Fonden's YIP OPTIC-AI project (grant n. 29344) and ERC CoG FRECOM, grant n. 771878.

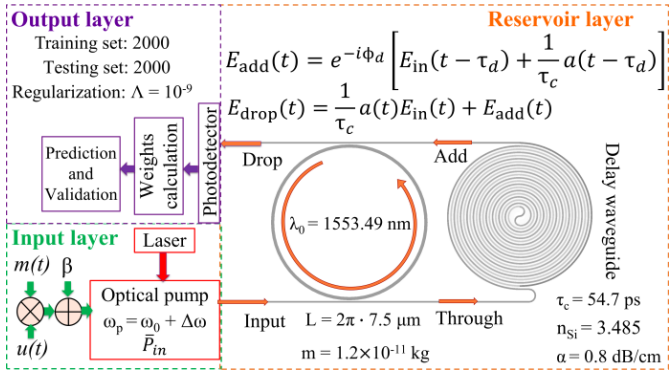

Fig. 1. TDRC scheme using a silicon MRR with delayed feedback.

delay $\tau_d$ to the feedback input, as proposed in [3]. We also consider the phase shift due to the delay waveguide propagation $\phi_d$.

The 1-GBd input $u(t)$ and mask $m(t)$ signals are generated as reported in [2] for the NARMA-10 task and an optimized bias $\beta=8.0$ added to their product so that the modulation index of the input signal remains less than <2%. The resulting signal modulates the pump with average input power $\bar{P}_{in}$. The RC operates at 50 GBd with 50 virtual nodes time-separated by $\tau_d = 0.5$ ns. We use ridge regression for the training, with the parameters specified in the output layer of Fig. 1. The MRR acts as the nonlinear node of the RC. The signal detected at the drop port of the MRR is used for training and testing the output. We evaluate the RC performance by calculating the normalized mean square error (NMSE), as defined in [2], between the target and the predicted output averaging over ten different seeds used to generate the target time series. In our simulations, we fix one of the parameters, $\tau_{FC}$ or $\tau_{th}$, while varying the other for a $\Delta\omega/2\pi$ range of ±200 GHz and a $\bar{P}_{in}$ range of –20 to +20 dBm.

## IV. RESULTS AND DISCUSSION

The NMSE as a function of $\Delta\omega$ and $\bar{P}_{in}$ (Fig. 2) shows the presence of a region with the worst performance (NMSE>1.0, in pink) of the RC. This is caused by a detrimental imbalance between the higher dimensionality given by the nonlinearities and the memory capacity required for a time-series prediction task. The memory capacity is decreased by the dominance of long-lasting nonlinear effects [3], i.e., when the TO effect dominates. In Fig. 2 (a-c) the NMSE is shown as a function of $\tau_{th}$. As the TO effect lasts longer, the frequency detuning and power range that is detrimental for the RC widens, yielding two narrow windows of low NMSE (blue and red detuning). The TO effect causes a red shift of the resonance and, when increasing $\tau_{th}$, the high-NMSE region extends towards negative frequency detuning. Therefore, controlling $\tau_{th}$ has the potential to enhance the performance of the RC. Alternatively, if we reduce $\tau_{FC}$ to the order of picoseconds (it has been shown in [7] that $\tau_{FC}$ can be reduced to ~12 ps), the region of low NMSE is extended as the shorter free-carrier relaxation time reduces the importance of the TO effect (Fig. 2 d-e). The largest frequency detuning range for desirable low-power operation can be found in Fig. 2 f), in which $\tau_{FC}$ is increased so that FCD effects become dominant. Hence, a blue shift of the resonance takes place, and the worst performance occurs in the positive detuning region. Thus, either reducing or increasing $\tau_{FC}$ may potentially stretch the region of low error. A minimum NMSE value of 0.0173 is reached for $\Delta\omega/2\pi$ = -50 GHz and $\bar{P}_{in}$= -5 dBm (red circle in Fig. 2 f), which is an improvement of performance for similar 50-node photonic reservoirs with similar speed [2].

## V. CONCLUSION

We have evaluated the performance of a silicon MRR-based RC architecture as a function of the relaxation times of free-carrier and thermal effects. We show that there is a strong influence of the properties of the microring waveguide on the region of frequency detuning and input power leading to low NMSE performance for NARMA-10 prediction. When numerically varying those physical parameters, we achieve an NMSE <0.05 with a complexity of 50 nodes.

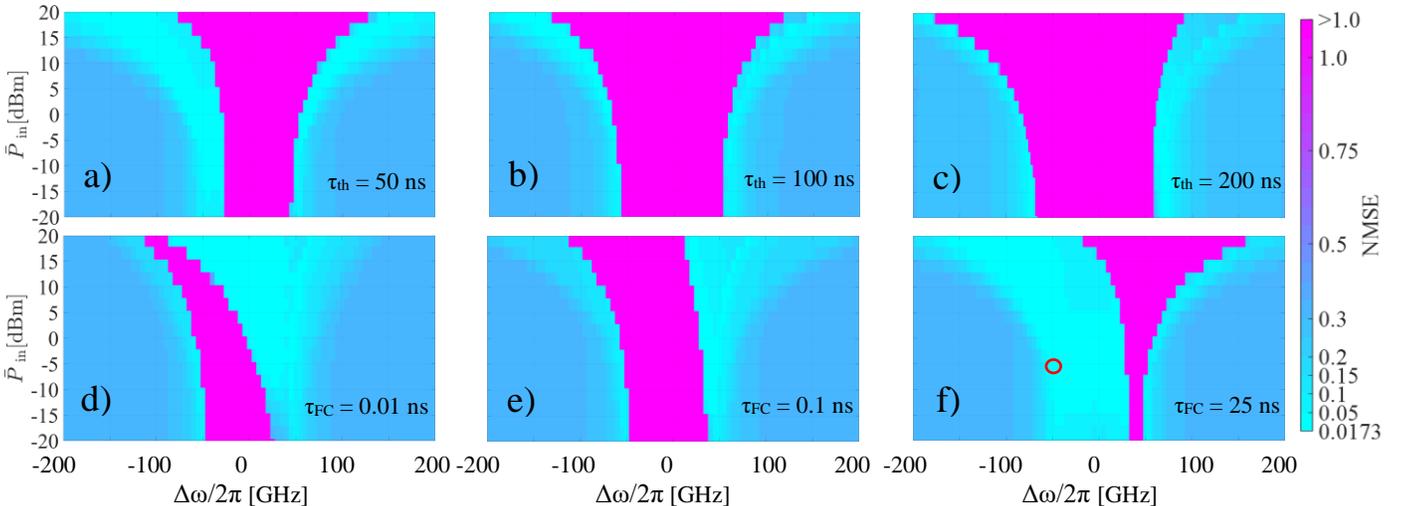

Fig. 2. NMSE performance of the RC for a) - c) increasing $\tau_{th}$ and $\tau_{FC}$ = 10 ns, and d) - f) increasing $\tau_{FC}$ and $\tau_{th}$ = 50 ns.